\documentclass[prd,nofootinbib,showpacs,preprint]{revtex4-1}
\usepackage[T1]{fontenc}
\usepackage{amsmath,amssymb}
\usepackage{epsfig}
\usepackage{graphicx}
\usepackage[usenames,dvipsnames]{color}
\usepackage{slashed}
\usepackage[colorlinks,citecolor=blue]{hyperref}
\usepackage{pdfpages}
\usepackage{color}
\begin{document}
\title{Light Sterile Neutrino and Dark Matter in Left-Right Symmetric Models Without Higgs Bidoublet}

\author{Debasish Borah}
\email{dborah@iitg.ernet.in}
\affiliation{Department of Physics, Indian Institute of Technology Guwahati, Assam 781039, India}
\begin{abstract}
We present a class of left-right symmetric models where Dirac as well as Majorana mass terms of neutrinos can arise at one-loop level in a scotogenic fashion: with dark matter particles going inside the loop. We show the possibility of naturally light right handed neutrinos that can have interesting implications at neutrinoless double beta decay experiments as well as cosmology. Apart from a stable dark matter candidate stabilised by a remnant $Z_2$ symmetry, one can also have a long lived keV sterile neutrino dark matter in these models. This class of models can have very different collider signatures compared to the conventional left-right models. 
\end{abstract}
\pacs{12.60.Fr,12.60.-i,14.60.Pq,14.60.St}
\maketitle

\section{Introduction}
Left-right symmetric model (LRSM) \cite{lrsm, lrsmpot} has been one of the most widely studied beyond standard model (BSM) scenarios in the last few decades. The model not only explains the sub-eV neutrino mass \cite{PDG, T2K, chooz, daya, reno, minos} naturally through seesaw mechanisms \cite{ti, tii0, tii, tiii} but also gives rise to low energy parity violations though spontaneous breaking of a parity preserving symmetry at high scale that can also be incorporated within grand unified theory groups like $SO(10)$ naturally. The minimal version of this model give rise to tiny neutrino masses through a combination of type I \cite{ti} and type II \cite{tii0, tii} seesaw mechanisms. Due to the structure of type I seesaw term, one usually has heavy right handed neutrinos in such a model, in order to generate sub-eV neutrino mass from type I seesaw. In TeV scale LRSM, one can have right handed neutrino mass $M_R \approx 1$ TeV (say) which can give sub-eV light neutrinos from type I seesaw if the Dirac neutrino Yukawa couplings should be fine tuned to $Y_{\nu} \leq 10^{-5.5}$. Similar fine-tuning is also involved in the type II seesaw term as we discuss in the next section. Such fine-tunings become more severe if we wish to keep one or two of the right handed neutrinos at very light scale, say between eV to keV. Such light sterile neutrinos could have very interesting implications for low energy neutrino oscillation experiments \cite{LSND1,miniboone,react,gall1,gall2}, neutrinoless double beta decay $(0\nu \beta \beta)$ \cite{kamland_zen, KamLAND-Zen:2016pfg,GERDA}, dark matter \cite{whitekev} and cosmology in general \cite{wmap9, Planck15}.

The LRSM also received significant attention after the recent hints from the large hadron collider (LHC) about the possibility of TeV scale new physics: the CMS $eejj$ excess \cite{CMSeejj}, the ATLAS diboson excess \cite{diboson} and more recently, the 750 GeV diphoton excess \cite{atlasconf,lhcrun2a,CMS:2015dxe}. The first two excesses around 2 TeV can be explained within a version of LRSM where the discrete left-right symmetry (to be introduced below) gets broken at a high energy scale \cite{Chang:1983fu} whereas the gauge symmetry of LRSM remains unbroken all the way down to the TeV scale. The recent 750 GeV diphoton excess can however, be explained within LRSM framework only when it is extended with additional vector like fermions \cite{LR750GeV1, LR750GeV2, LR750GeV3, LR750GeV4} or with fields having high $SU(2)$ dimensions \cite{LR750GeV5}. The vector like fermions, as discussed in these works, can serve three purposes: (i) to explain the large diphoton cross section at 750 GeV, (ii) to provide masses to all the fermions of the standard model (SM) through a universal seesaw mechanism \cite{VLQlr, univSeesawLR} and (ii) to assist in gauge coupling unification \cite{LR750GeV3, LR750GeV4}. This model can also have interesting implications for cosmology \cite{gulr} and $0\nu \beta \beta$ \cite{lr0nu2beta}. 

In the present work, we study a class of left-right symmetric models to see the possibility of having light sterile neutrinos at a scale much below the scale of left-right symmetry. Typically, the right handed neutrinos in LRSM have masses around the scale of left-right symmetry. Thus, in TeV scale LRSM, one usually have right handed neutrinos in the GeV to TeV range. If we fine tune the Yukawa couplings to be as small as electron Yukawa, then we can get right handed neutrinos of the order of tens of MeV. We therefore study some non-minimal versions of LRSM where one can have light right handed neutrinos in the eV-keV regime even with order one Yukawa couplings. Starting with the LRSM having universal seesaw for all fermions, we explore a few other possibilities: (i) light neutrinos as Dirac fermions, (ii) both left and right handed neutrinos acquire Majorana mass terms through radiative type II seesaw and (iii) left handed neutrinos acquire masses through radiative type I seesaw. In all of these possibilities, we consider additional discrete symmetries to stabilise the particles going inside the loops so that the lightest neutral particle among them can be a cold dark matter (CDM) candidate. This follows the basic idea of scotogenic models first proposed by \cite{m06}. Apart from the stable cold dark matter candidate, these models can also have a long lived keV scale right handed neutrino that can be a warm dark matter (WDM) candidate. Such keV scale sterile neutrino dark matter within minimal LRSM was discussed by \cite{wdmlr1, wdmlr2} and a radiative neutrino mass model with both cold and warm dark matter components have been proposed recently by \cite{abm16}. We also discuss the possible implications of these scenarios in $0\nu \beta \beta$, collider experiments and cosmology.

In all the models we discuss in this work, there exists additional vector like fermions which are singlets under the left-right gauge symmetry. Since they are singlets, their bare mass terms can be written in the Lagrangian and the symmetry of the theory does not restrict their masses to the TeV scale. In a generic theory, one expects these bare mass terms to be close to the grand unified theory (GUT) scale or the Planck scale. This is in contrast with the minimal LRSM where all the fermions acquire masses as a result of spontaneous gauge symmetry breaking. One can however introduce a discrete $Z_2$ symmetry forbidding the bare mass terms of vector like fermions in a way that was shown by the authors of \cite{LR750GeV1}. Another singlet scalar with appropriate $Z_2$ charge can be introduced in such a way to give rise to a Yukawa term for the vector like fermions. This singlet scalar can then acquire a non-zero vacuum expectation value (vev) and explain the masses of the vector like fermions in a dynamical manner. Also, the fermion content of minimal LRSM can be accommodated within the spinor representation $\textbf{16}$ of $SO(10)$ GUT models which is not possible in the present class of models. Although the present class of models can give rise to gauge coupling unification \cite{LR750GeV3, LR750GeV4}, the GUT group should be bigger than the minimal $SO(10)$ to accommodate the vector like fermions.

This article is organised as follows. In section \ref{sec1}, we briefly discuss the minimal LRSM and then discuss the LRSM with universal seesaw in section \ref{sec2}. In section \ref{sec3} we discuss different possible versions of scotogenic LRSM where neutrinos acquire masses at one loop level with dark matter particles going inside the loops. In section \ref{sec4}, we discuss the possibilities of light sterile or right handed neutrinos in these models followed by their implications in $0\nu \beta \beta$ and colliders in section \ref{sec5} and \ref{sec6} respectively. We then comment upon cosmological implications of these scenarios in section \ref{sec7} and finally conclude in section \ref{sec8}.

\section{Minimal Left-Right Symmetric Model (MSLRM)}
\label{sec1}
Left-Right Symmetric Model \cite{lrsm,lrsmpot} is one of the best motivated BSM frameworks where the gauge symmetry of the electroweak theory is extended to $SU(3)_c \times SU(2)_L \times SU(2)_R \times U(1)_{B-L}$. The right handed fermions are doublets under $SU(2)_R$ similar to the way left handed fermions transform as doublets under $SU(2)_L$. The requirement of an anomaly free $U(1)_{B-L}$ makes the presence of right handed neutrinos a necessity rather than a choice. To allow Dirac Yukawa couplings between $SU(2)_{L,R}$ doublet fermions, the Higgs field has to transform as a bidoublet under $SU(2)_{L,R}$. In order to break the gauge symmetry of the model to that of the SM spontaneously, scalar triplet fields with non-zero $U(1)_{B-L}$ charges are introduced, which also give Majorana masses to the left and right handed neutrinos.

The fermion content of the minimal LRSM is
\begin{equation}
Q_L=
\left(\begin{array}{c}
\ u_L \\
\ d_L
\end{array}\right)
\sim (3,2,1,\frac{1}{3}),\hspace*{0.8cm}
Q_R=
\left(\begin{array}{c}
\ u_R \\
\ d_R
\end{array}\right)
\sim (3^*,1,2,\frac{1}{3}),\nonumber 
\end{equation}
\begin{equation}
\ell_L =
\left(\begin{array}{c}
\ \nu_L \\
\ e_L
\end{array}\right)
\sim (1,2,1,-1), \quad
\ell_R=
\left(\begin{array}{c}
\ \nu_R \\
\ e_R
\end{array}\right)
\sim (1,1,2,-1) \nonumber
\end{equation}
Similarly, the Higgs content of the minimal LRSM is
\begin{equation}
\Phi=
\left(\begin{array}{cc}
\ \phi^0_{11} & \phi^+_{11} \\
\ \phi^-_{12} & \phi^0_{12}
\end{array}\right)
\sim (1,2,2,0)
\nonumber 
\end{equation}
\begin{equation}
\Delta_L =
\left(\begin{array}{cc}
\ \delta^+_L/\surd 2 & \delta^{++}_L \\
\ \delta^0_L & -\delta^+_L/\surd 2
\end{array}\right)
\sim (1,3,1,2), \hspace*{0.2cm}
\Delta_R =
\left(\begin{array}{cc}
\ \delta^+_R/\surd 2 & \delta^{++}_R \\
\ \delta^0_R & -\delta^+_R/\surd 2
\end{array}\right)
\sim (1,1,3,2) \nonumber
\end{equation}
where the numbers in brackets denote the transformations of the fields under the gauge group $SU(3)_c\times SU(2)_L\times SU(2)_R \times U(1)_{B-L}$. During the spontaneous symmetry breaking of LRSM gauge group down to the SM gauge group, the neutral component of the Higgs triplet $\Delta_R$ acquires a non-zero vev after which the neutral components of Higgs bidoublet $\Phi$ acquire non-zero vev's to break the SM gauge symmetry into the $U(1)$ of electromagnetism. This symmetry breaking chain can be denoted as:
$$ SU(2)_L \times SU(2)_R \times U(1)_{B-L} \quad \underrightarrow{\langle
\Delta_R \rangle} \quad SU(2)_L\times U(1)_Y \quad \underrightarrow{\langle \Phi \rangle} \quad U(1)_{em}$$
Denoting the vev of the two neutral components of the bidoublet as $k_1, k_2$ and that of triplets $\Delta_{L, R}$ as $v_{L, R}$ and considering $g_L=g_R$,  $k_2 \sim v_L \approx 0$ and $v_R \gg k_1$, the approximate expressions for gauge boson masses after symmetry breaking can be written as 
$$ M^2_{W_L} = \frac{g^2}{4} k^2_1, \;\;\; M^2_{W_R} = \frac{g^2}{2}v^2_R $$
$$ M^2_{Z_L} =  \frac{g^2 k^2_1}{4\cos^2{\theta_w}} \left ( 1-\frac{\cos^2{2\theta_w}}{2\cos^4{\theta_w}}\frac{k^2_1}{v^2_R} \right), \;\;\; M^2_{Z_R} = \frac{g^2 v^2_R \cos^2{\theta_w}}{\cos{2\theta_w}} $$
where $\theta_w$ is the Weinberg angle.

The relevant Yukawa couplings for fermion masses can be written as 
\begin{eqnarray}
{\cal L}^{II}_\nu &=& y_{ij} \bar{\ell}_{iL} \Phi \ell_{jR}+ y^\prime_{ij} \bar{\ell}_{iL}
\tilde{\Phi} \ell_{jR} +Y_{ij} \bar{q}_{iL} \Phi q_{jR}+ Y^\prime_{ij} \bar{q}_{iL}
\tilde{\Phi} q_{jR} +\text{h.c.}
\nonumber \\
&+& f_{ij}\ \left(\ell_{iR}^T \ C \ i \sigma_2 \Delta_R \ell_{jR}+
(R \leftrightarrow L)\right)+\text{h.c.}
\label{treeY}
\end{eqnarray}
where $\tilde{\Phi} = \tau_2 \Phi^* \tau_2$. In the above Yukawa Lagrangian, the indices $i, j = 1, 2, 3$ correspond to the three generations of fermions. The Majorana Yukawa couplings $f$ is same for both left and right handed neutrinos
because of the in built left-right symmetry $(f_L = f_R)$. These couplings $f$ give rise to the Majorana mass terms of both left handed and right handed neutrinos after the triplet Higgs fields $\Delta_{L,R}$ acquire non-zero vev. Although it is the $\Delta_R$ field which gets a vev at high scale breaking the left-right symmetry, the subsequent electroweak symmetry breaking induces a non-zero vev to the left handed counterpart. The induced vev for the left-handed triplet $v_{L}$ can be shown for generic LRSM to be
$$v_{L}=\gamma \frac{M^{2}_{W_L}}{v_{R}}$$
with $M_{W_L}\sim 80.4$ GeV being the weak boson mass such that 
$$ |v_{L}|<<M_{W_L}<<|v_{R}| $$ 
In general $\gamma$ is a function of various couplings in the scalar potential of generic LRSM. Using the results from Deshpande et al., \cite{lrsmpot}, $\gamma$ is given by
\begin{equation}
\gamma = \frac{\beta_2 k^2_1+\beta_1 k_1 k_2 + \beta_3  k^2_2}{(2\rho_1-\rho_3)(k^2_1+k^2_2)}
\label{eq:gammaLR}
\end{equation}
where $\beta, \rho$ are dimensionless parameters of the scalar potential. Without any fine tuning $\gamma$ is expected to be of the order unity ($\gamma\sim 1$). However, for TeV scale type I+II seesaw, $\gamma$ has to be fine-tuned as we discuss later. The type II seesaw formula for light neutrino masses can now be expressed as
\begin{equation}
M_{\nu}=\gamma (M_{W_L}/v_{R})^{2}M_{RR}-m_{LR}M^{-1}_{RR}m^{T}_{LR}
\label{type2}
\end{equation}
If we consider the first term on the right hand side of the above expression, one can make an estimate of neutrino mass for TeV scale LRSM. Considering $v_R \sim 6$ TeV, the type II seesaw term will be of the order of light neutrino mass $M_{\nu} \sim 0.1$ eV if 
$$ \gamma \approx \frac{5.6 \times 10^{-7}}{M_R} $$
where $M_R$ is the right handed neutrino mass. Thus, for TeV scale right handed neutrino masses, the dimensionless parameter $\gamma$ is fine tuned to the level of $10^{-10}-10^{-9}$ in order to get correct order of neutrino masses. This will involve unnatural fine tuning of the scalar potential parameters appearing in the expression for $\gamma$ given in \eqref{eq:gammaLR}. Similar but slightly less fine tuning is involved in the type I seesaw term for TeV scale $M_{RR}$. The Dirac Yukawa couplings should be fine tuned to around $10^{-6}-10^{-5}$ in order to get light neutrino mass of order $0.1$ eV.

\section{LRSM with universal seesaw}
\label{sec2}
In minimal LRSM with Higgs bidoublet and Higgs triplets discussed above, the light neutrino masses arise at tree level. If the triplets are replaced by Higgs doublets, then additional fermions are required for seesaw mechanisms. In the absence of them, neutrinos have only a Dirac mass term due to the bidoublet, which typically is of the order of charged lepton masses. Thus, the tree level Dirac mass term of neutrinos can be prevented only in the absence of Higgs bidoublet. Indeed, LRSM without Higgs bidoublet has found some attention in the literature \cite{VLQlr, univSeesawLR}. This model was studied later in the context of cosmology \cite{gulr} and neutrinoless double beta decay \cite{lr0nu2beta}. Very recently this model was also studied in the context of 750 GeV di-photon excess at LHC \cite{lhcrun2a,atlasconf,CMS:2015dxe} by several authors \cite{LR750GeV1, LR750GeV2, LR750GeV3, LR750GeV4}. The model has additional vector like fermions in order to generate all the fermion masses from universal seesaw. The fermion content of the LRSM with universal seesaw is the extension of the MLRSM fermion content by the following vector like fermions 
$$ U_{L} (3, 1, 1, \frac{4}{3}), \;\; U_{R} (3^*, 1, 1, \frac{4}{3})\;\; D_{L} (3, 1, 1, -\frac{2}{3}), \;\;D_{R} (3^*, 1, 1, -\frac{2}{3}) $$
$$ E_{L,R} (1,1,1, -2), \;\; N_{L,R} (1,1,1,0) $$
per generation of quarks and leptons. Instead of Higgs bidoublet, this model has a pair of Higgs doublets 
$$ H_L (1, 2, 1, -1), \;\;H_R (1,1,2,-1)$$

The non-zero vev of the neutral component of $H_R$ also breaks the $SU(2)_R\times U(1)_{B-L}$ symmetry of the model into $U(1)_Y$ of the standard model. The left handed Higgs doublet can acquire a non-zero vev $v_L$ at a lower energy to induce electroweak symmetry breaking. However, the left-right symmetry of the theory forces one to have the same vev for both $H_L$ and $H_R$ that is, $v_L=v_R$ which is unacceptable from phenomenological point of view. To decouple these two symmetry breaking scales, the extra singlet scalar $\sigma$ is introduced into the model. This field is odd under the discrete left-right symmetry and hence couple to the two scalar doublets with a opposite sign. After this singlet acquires a non-zero vev at high scale, this generates a difference between the effective mass squared of $H_L$ and $H_R$ which ultimately decouples the symmetry breaking scales. 

Due to the absence of the usual bidoublet, the left and right handed fermion doublets of the MSLRM can not directly couple to each other. However, they can couple to the scalar fields $H_{L,R}$ via the additional vector like fermions. 
\begin{align}
\mathcal{L} & \supset Y_U (\bar{Q_L} H_L U_L+\bar{Q_R} H_R U_R) + Y_D (\bar{Q_L} H^{\dagger}_L D_L+\bar{Q_R} H^{\dagger}_R D_R) +M_U \bar{U_L} U_R+ M_D \bar{D_L} D_R\nonumber \\
& +Y_E (\bar{\ell_L} H^{\dagger}_L E_L+\bar{\ell_R} H^{\dagger}_R E_R) +Y_\nu (\bar{\ell_L} H_L N_L+\bar{\ell_R} H_R N_R)+M_E \bar{E_L} E_R+ M^D_N \bar{N_L} N_R \nonumber \\
& +\frac{1}{2} M^M_N (N_L N_L + N_R N_R)+\text{h.c.}
\end{align}
After integrating out the heavy fermions, the charged fermions of the standard model develop Yukawa couplings to the scalar doublet $H_L$ as follows
$$ y_u = Y_U \frac{v_R}{M_U} Y^T_U, \;\;y_d = Y_D \frac{v_R}{M_D} Y^T_D, \;\;y_e = Y_E \frac{v_R}{M_E} Y^T_E $$
where $v_R$ is the vev of the neutral component of $H_R$. The apparent seesaw then can explain the observed mass hierarchies among the three generations of fermions.  The neutrino mass arises in a more complicated seesaw due to additional Majorana mass terms for $N_L, N_R$ shown above. As shown in \cite{gulr}, the neutrino mass matrix in the basis $(\nu_L, \nu_R)$ can have three independent terms
$$ m_L = -Y_\nu \frac{v_L}{M^M_N} Y^T_\nu v_L $$
$$ m_R = -Y_\nu \frac{v_R}{M^M_N} Y^T_\nu v_R $$
$$ m_D = Y_\nu \frac{1}{M^M_N} (M^D_N)^T \frac{1}{M^M_N} Y^{\dagger}_\nu v_L v_R $$
after integrating out the heavy neutral fermions $N_{L,R}$.  This mass matrix can further be diagonalised to get the type I seesaw formula for neutrino mass which however, is suppressed by $\left (\frac{M^D_N}{M^M_N} \right)^2$ compared to $m_L$. Thus, the left and right handed neutrinos have Majorana mass terms which are related by 
\begin{equation}
m_R = m_L \frac{v^2_R}{v^2_L}
\end{equation}
If we have a TeV scale left-right symmetry breaking say $v_R \approx 6$ TeV, then the above relation says $m_R  \approx 595 m_L$. Thus for $m_L \leq 0.1$ eV, the right handed neutrino masses are $m_R \leq 59.5$ eV. Such light right handed neutrinos will have very interesting implications in cosmology, collider as well as neutrinoless double beta decay experiments, when the LR gauge boson masses are around the TeV corner. One can however, keep them hidden in neutrinoless double beta decay experiments by tuning the $ee$ element of the right handed neutrino mass matrix to zero. On the other hand, direct search experiments like the LHC search for heavy neutrinos and $W_R$ bosons at 8 TeV centre of mass energy already excludes a wide range of heavy neutrino masses \cite{CMSNRWR}. However, as can be seen from the $m_R-M_{W_R}$ exclusion plot in \cite{CMSNRWR}, the experimental sensitivity is not so good for the low mass regime of right handed neutrinos. Also, a right handed neutrino in the eV-keV regime will not give rise to the same sign dilepton plus two jets signatures which are being looked at by LHC experiments to put bounds on $m_R-M_{W_R}$ parameter space \cite{CMSNRWR}. Such light right handed neutrinos will also induce non-unitarity to leptonic mixing matrix of the order 
\begin{equation}
\eta \approx \frac{m_D}{m_R} \approx \frac{M^D_N}{M^M_N} \frac{v_L}{v_R}
\label{nonunitary}
\end{equation}
which will again constrain $\frac{M^D_N}{M^M_N}$ to a small number for TeV scale $v_R$. If we want to keep the right handed neutrino mass at keV scale, the left-right symmetry breaking scale has to be $v_R \sim 25$ TeV, beyond the reach of LHC.

\section{Scotogenic LRSM}
\label{sec3}

Since the TeV scale LRSM with universal seesaw for all fermions generate three ultra light right handed neutrinos, which may be in conflict with experiments from colliders to cosmology, one can find suitable alternative to this. One such immediate remedy is to consider universal seesaw for charged fermions and radiative masses for neutrinos. One such possibility was discussed very recently in \cite{radiativeLR} where neutrino mass was generated at two-loop level along with a bonus of having a long lived keV neutrino dark matter. Here we consider another possibility of having radiative neutrino mass where the particles running inside the loop can be stable dark matter candidates. We consider two possibilities where light neutrinos can be either of Dirac type or of Majorana type.

\subsection{Radiative Dirac Neutrino Mass}
\begin{figure}[!h]
\centering
\epsfig{file=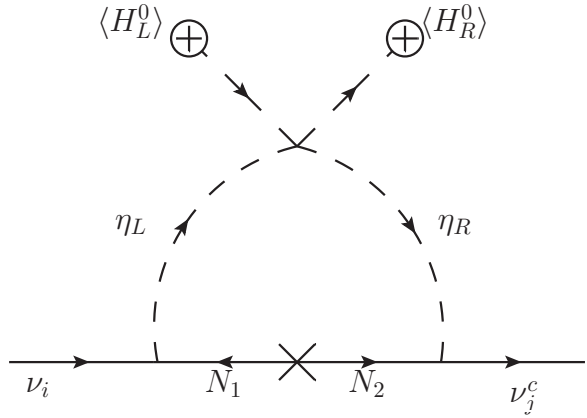,width=0.50\textwidth,clip=}
\caption{One-loop Dirac neutrino mass for the particle content shown in table \ref{tab:data1}}
\label{fig1}
\end{figure}

Let us consider a version of LRSM with universal seesaw for all the charged fermions but zero tree level masses for neutrinos. This corresponds to the model discussed above but without the neutral fermion pairs $N_{L,R}$. Instead of that, consider the addition of following particles into the model with their transformations under the LRSM gauge group as well as an additional symmetry forbidding tree level neutrino mass and guaranteeing the stability of dark matter. These new particle sector consists of scalar doublets $\eta_{L,R}$ and fermion singlets $N$. The additional symmetry can either be discrete $Z_N$ or a continuous one like $U(1)_X$. We show one simple example of $Z_4$  in this subsection. The fields taking part in radiative neutrino mass generation are shown in table \ref{tab:data1}. Since the Higgs doublets $H_{L,R}$ have trivial transformations under the additional $Z_4$, the usual mass terms for charged fermions are allowed. All other particles in LRSM with universal seesaw for charged fermions do not transform under this additional symmetry. Such a structure of new particles and symmetry will give rise to the one-loop Dirac neutrino mass shown in figure \ref{fig1}. One loop Dirac neutrino mass in a scotogenic fashion was first proposed in \cite{mafarzan12} within a model where the SM gauge symmetry was extended by $U(1)_{B-L}$ and additional $Z_2$ discrete symmetries.

\begin{table}
\begin{center}
\begin{tabular}{|c|c|c|}
\hline
Particles & $SU(3)_c \times SU(2)_L \times SU(2)_R \times U(1)_{B-L} \times P $  & $n_p, P\equiv Z_4$  \\
\hline
$H_L$ & $(1,2,1,-1, n_p)$ & $1$  \\
$H_R$ & $(1,1,2,-1, n_p)$ & $1$ \\
$\eta_L$ & $(1,2,1,-1, n_p)$ & $i$  \\
$\eta_R$ & $(1,1,2,-1, n_p)$ & $i$ \\
$ N_1 $ & $(1,1,1,0,n_p)$ & $i$\\
$ N_2 $ & $(1,1,1,0,n_p)$ & $-i$\\
\hline
\end{tabular}
\end{center}
\caption{Particle content for radiative Dirac neutrino masses}
\label{tab:data1}
\end{table}


\begin{table}
\begin{center}
\begin{tabular}{|c|c|c|}
\hline
Particles & $SU(3)_c \times SU(2)_L \times SU(2)_R \times U(1)_{B-L} \times P $  & $n_p, P\equiv Z_2$  \\
\hline
$\eta_L$ & $(1,2,1,-1, n_p)$ & $-1$  \\
$\eta_R$ & $(1,1,2,-1, n_p)$ & $-1$ \\
$ N_1 $ & $(1,1,1,0,n_p)$ & $-1$\\
\hline
\end{tabular}
\end{center}
\caption{Minimum additional fields responsible for radiative Majorana neutrino masses}
\label{tab:data2}
\end{table}
\begin{table}
\begin{center}
\begin{tabular}{|c|c|c|}
\hline
Particles & $SU(3)_c \times SU(2)_L \times SU(2)_R \times U(1)_{B-L} \times P $  & $n_p, P\equiv Z_4$  \\
\hline
$\eta_L$ & $(1,2,1,-1, n_p)$ & $i$  \\
$\eta_R$ & $(1,1,2,-1, n_p)$ & $i$ \\
$S$ & $(1,1,1,0, n_p)$ & $-1$ \\
$\Delta_L$ & $(1,3,1,2, n_p)$ & $-1$ \\
$\Delta_R$ & $(1,1,3,2, n_p)$ & $-1$ \\
$ N_1 $ & $(1,1,1,0,n_p)$ & $i$\\
\hline
\end{tabular}
\end{center}
\caption{Radiative type II seesaw for Majorana neutrino masses}
\label{tab:data3}
\end{table}

\subsection{Radiative Majorana Neutrino Mass}
\begin{figure}[!h]
\centering
\epsfig{file=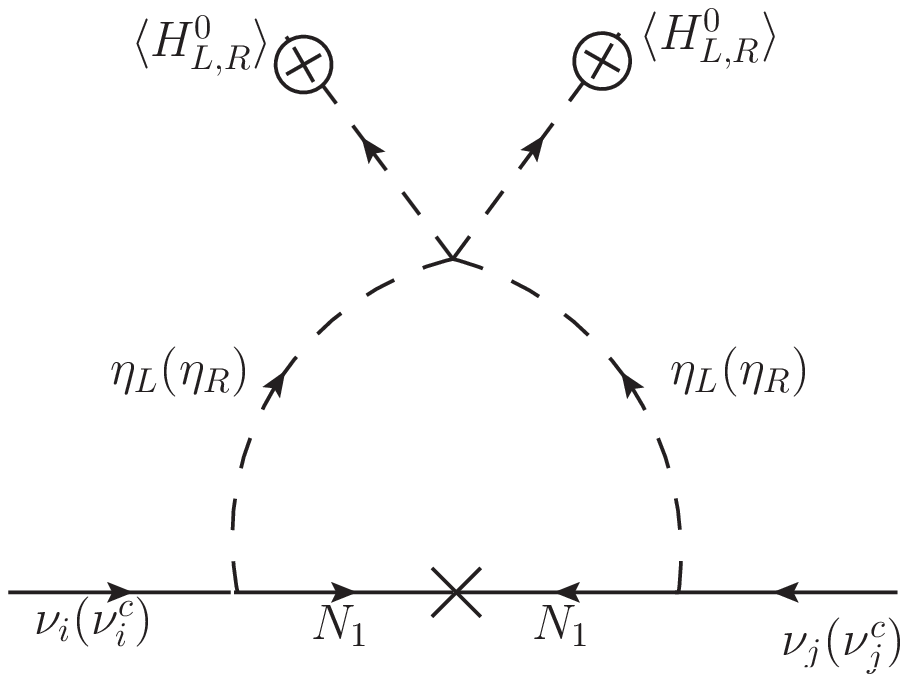,width=0.50\textwidth,clip=}
\caption{One-loop Majorana neutrino mass}
\label{fig1a}
\end{figure}

One-loop Majorana masses of both left and right handed neutrinos shown in figure \ref{fig1a} can be obtained from the additional fields shown in table \ref{tab:data2}. Since the couplings in left and right sectors are identical due to left-right symmetry, such a diagram will generate identical Majorana masses for left and right handed neutrinos. Thus, the simple implementation of the scotogenic idea within LRSM also generates very light right handed Majorana neutrinos similar to the LRSM with universal seesaw model with TeV scale $W_R$ bosons. 

One can break the degeneracy between left and right handed neutrino masses by introducing a pair of scalar triplets $\Delta_{L,R}$ similar to the MLRSM. However, since tree level type II seesaw suffers from fine tuning issues in TeV scale MLRSM as discussed earlier, here we consider two alternatives: (a) Allow radiative type II seesaw for both left and right handed neutrinos and (b) Allow tree level coupling of triplets to neutrinos but forbid the induced vev of left handed triplet; such that the light neutrino mass arises as a consequence of type I seesaw mechanism where the Dirac mass term arises radiatively.

\begin{figure}[!h]
\centering
\begin{tabular}{cc}
\epsfig{file=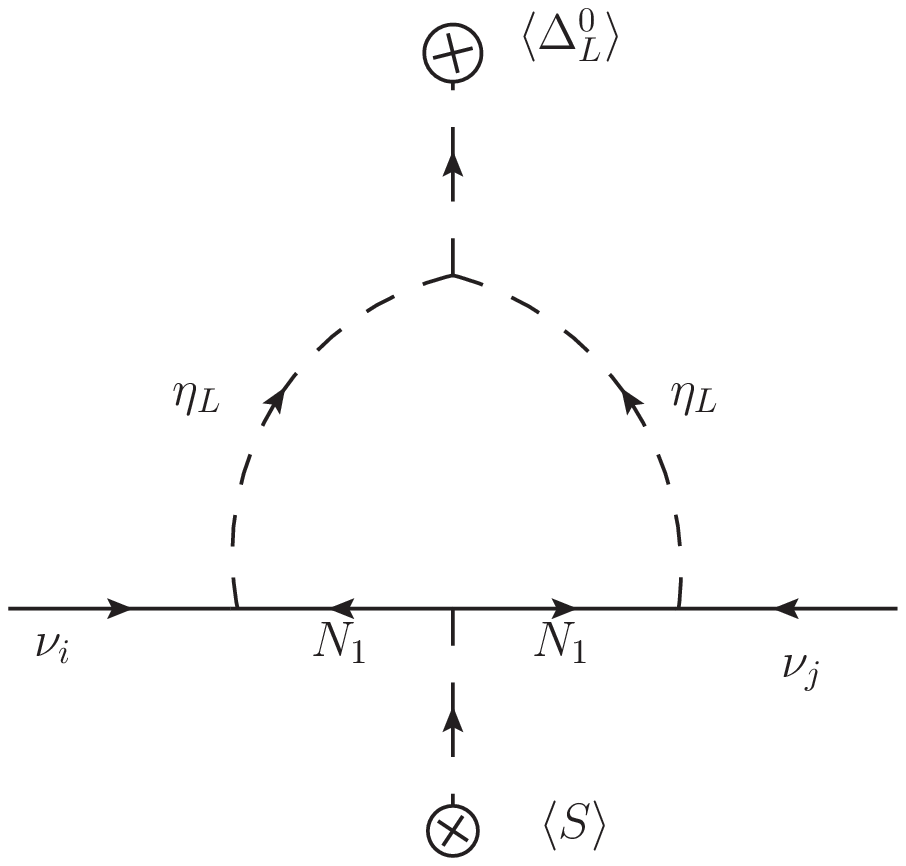,width=0.50\textwidth,clip=} & 
\epsfig{file=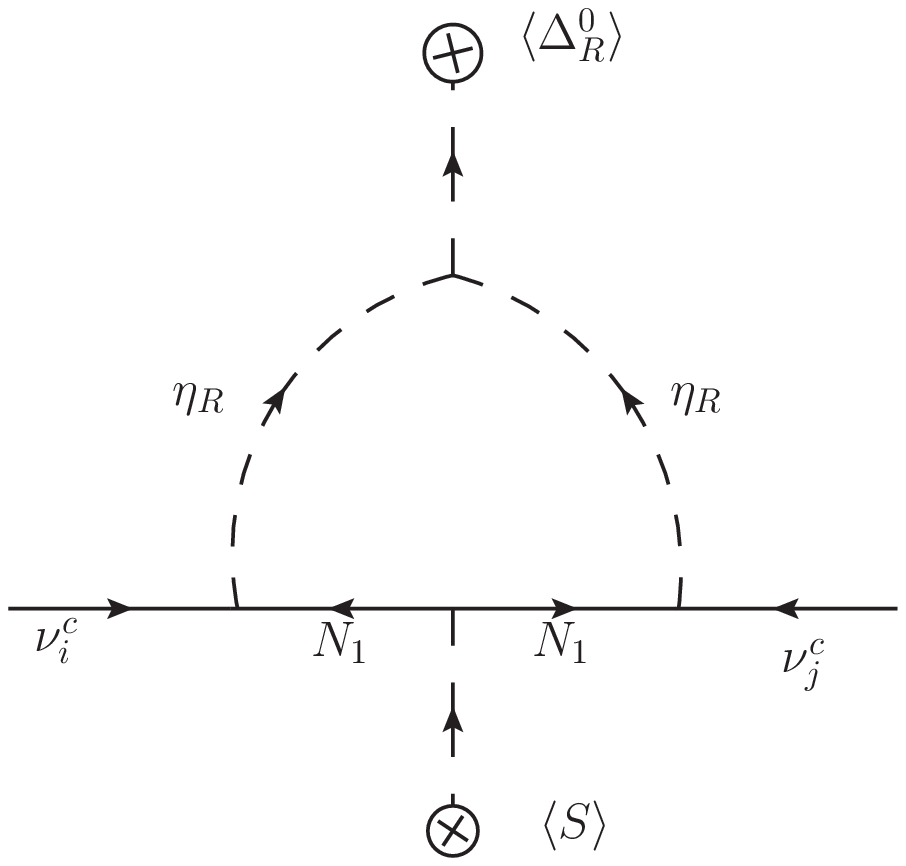,width=0.50\textwidth,clip=}
\end{tabular}
\caption{One-loop Majorana neutrino mass}
\label{fig2}
\end{figure}

\textit{Model (a)}: Radiative type II seesaw in LRSM can be implemented with the additional particle content shown in table \ref{tab:data3}. The relevant one-loop diagrams are shown in figure \ref{fig2}. Since the neutral components of $\Delta_L$ and $\Delta_R$ acquire different vev, one can in general generate different Majorana masses for left and right handed neutrinos. Here also, the ratio of left and right handed neutrino masses is proportional to $v_L/v_R$ similar to the type II seesaw in MLRSM. Thus for TeV scale $v_R$, one has to do fine tuning in the scalar potential parameters in order to generate a tiny $v_L$.

\textit{Model (b)}: In the MLRSM both the scalar triplets couple to leptons such that right handed neutrinos acquire masses from $\Delta_R$ vev whereas the left handed ones acquire type II seesaw mass from left scalar triplet. The left scalar triplet acquires an induced vev after EWSB by virtue of its couplings to the electroweak bidoublet. In LRSM with Higgs doublets also, the triplets can couple to the Higgs doublets and the left triplet can acquire a non-zero vev after EWSB. For TeV scale $v_R$, we usually get a large $v_L$ and substantial amount of fine-tuning is required in order to keep the type II seesaw term at sub-eV scale. We can however, forbid the doublet-triplet scalar coupling term by introducing additional symmetries, such that the left handed triplet does not acquire any induced vev after the EWSB. This can be done with an additional $U(1)_X$ gauge symmetry shown in table \ref{tab:data4} under which the triplet scalars $\Delta_{L,R}$ do not transform. However, due to non-trivial charges of $H_{L,R}$ under $U(1)_X$ one can not write down terms like $H^T_L \Delta_L H_L$ in the scalar potential. This prevents the induced vev of $\Delta_L$ and hence the absence of type II seesaw. In this case, neutrinos acquire a Dirac mass term at one-loop level as seen from figure \ref{fig1} and right handed neutrinos acquire a Majorana mass term from the vev of $\Delta_R$. The effective light neutrino masses will arise from a seesaw between Dirac neutrino mass and right handed Majorana neutrino mass, as in the type I seesaw mechanism.

\begin{table}
\begin{center}
\begin{tabular}{|c|c|c|}
\hline
Particles & $SU(3)_c \times SU(2)_L \times SU(2)_R \times U(1)_{B-L} \times P $  & $n_p, P\equiv U(1)_X$  \\
\hline
$H_L$ & $(1,2,1,-1, n_p)$ & $n_h$  \\
$H_R$ & $(1,1,2,-1, n_p)$ & $n_h$ \\
$\eta_L$ & $(1,2,1,-1, n_p)$ & $n$  \\
$\eta_R$ & $(1,1,2,-1, n_p)$ & $n$ \\
$\Delta_L$ & $(1,3,1,2, n_p)$ & $0$ \\
$\Delta_R$ & $(1,1,3,2, n_p)$ & $0$ \\
$ N_1 $ & $(1,1,1,0,n_p)$ & $n$\\
$ N_2 $ & $(1,1,1,0,n_p)$ & $-n$\\
\hline
\end{tabular}
\end{center}
\caption{Radiative type I seesaw for Majorana neutrino masses}
\label{tab:data4}
\end{table}

\section{Possibility of light sterile neutrino}
\label{sec4}
In the models discussed above, one can easily have light sterile neutrinos at eV-keV scale depending on the left-right symmetry breaking scale $v_R$. In MLRSM also, it is possible to achieve a keV scale sterile neutrino which can be warm dark matter, by suitable fine-tuning in the neutrino mass expression from type I seesaw. Such a possibility was discussed in \cite{wdmlr1, wdmlr2}. For a TeV scale MLRSM, this will involve fine tunings at the level of $10^{-9}$ in the Yukawa couplings appearing in the mass term of right handed neutrinos. Similar fine tunings will also be required in Dirac Yukawa couplings of neutrinos in order to arrive at sub-eV or lower light neutrino masses from type I seesaw. In the LRSM with universal seesaw discussed in section \ref{sec2}, it is easier to achieve light sterile neutrinos without much fine-tuning. Since the right and left handed neutrino masses are different by a factor $\frac{v^2_R}{v^2_L}$ where $v_L = 246$ GeV is the electroweak vev, one can easily adjust the scale $v_R$ in order to have the desired sterile neutrino spectrum. For example, if we want the heaviest sterile neutrino mass to be around a few keV, suitable for warm dark matter candidates, we need to push the left-right breaking scale $v_R$ to beyond 60 TeV, far beyond the reach of current experiments. Thus, in TeV version of this model, there is no dark matter candidate either in cold dark matter or warm dark matter sectors. However, there can be light sterile neutrinos around eV scale in the TeV scale version of this model, which can play interesting role in neutrino oscillation and $0\nu \beta \beta$ experiments.

In the scotogenic model with purely Dirac neutrinos, the light neutrinos neutrinos are Dirac fermions similar to the charged fermions of standard model. Such a scenario does not have any extra sterile species between eV-keV scale. Thus, the dark sector consists of only cold dark matter in the form of inert Higgs $\eta_{L,R}$ or inert singlet fermions $N_1$. For the simplest scotogenic extension of LRSM with light Majorana neutrinos, the relevant extra particle content of which is shown in table \ref{tab:data2}, the light neutrino sector consists of three active and three sterile neutrinos with identical masses. Since similar diagrams with mixed $\eta_L, \eta_R$ internal lines in figure \ref{fig1a} will give rise to active-sterile mixing, such a scenario will be ruled out constraints from neutrino oscillation experiments. In particular, \cite{globalfit} does a global analysis of the available data and provides constraints on the active-sterile mixings which rules out sub-eV sterile neutrinos with identical masses as active neutrinos and large mixing.

In the scotogenic model with radiative type II seesaw, one can have light sterile neutrinos depending on the vev ratio of $\Delta^0_L, \Delta^0_R$. Using the expression from \cite{m06} of one-loop neutrino mass 
\begin{equation}
(m_{\nu})_{ij} = \sum_k \frac{h_{ik}h_{jk} M_{k}}{16 \pi^2} \left ( \frac{m^2_R}{m^2_R-M^2_k} \text{ln} \frac{m^2_R}{M^2_k}-\frac{m^2_I}{m^2_I-M^2_k} \text{ln} \frac{m^2_I}{M^2_k} \right)
\end{equation}
where $h_{ij}$ are Yukawa couplings for the terms $\bar{L} \eta_L N_1$ in the Lagrangian, $m^2_{R,I}$ are the masses of scalar and pseudoscalar part of $\eta^0_{L}$ and $M_k$ the mass of singlet fermion $N$ in the internal line. In the simplest assumption of $M^2_k \approx  m^2_0 = (m^2_R+m^2_I)/2$ we can write down the expression for light neutrino mass as
\begin{equation}
(m_{\nu})_{ij} = \frac{\mu v_L}{16 \pi^2} \sum_k \frac{h_{ik}h_{jk}}{ M_{k}}
\end{equation}
where we have replaced the mass difference $m^2_R-m^2_I = 2 \mu v_L$. Such a mass difference arises from the terms $\frac{1}{2} ( \mu \eta^T_L \Delta^{\dagger}_L \eta_L + \text{h.c.})$ in the scalar potential. The expression for right handed neutrino mass will be similar with $v_L$ being replaced by $v_R$. Taking the maximum possible value of $v_L$ which is around 5 GeV, allowed by $\rho$ parameter constraints \cite{PDG14} one can have the heaviest right handed neutrino mass at around $120$ eV, if we fix the LRSM scale $v_R$ to be 6 TeV. For a few keV right handed neutrinos with TeV scale LRSM, one needs to keep $v_L$ at a few tens of MeV. Depending on the hierarchy of light neutrinos, one can also have one of the sterile neutrinos at sub-eV scale, which may have interesting signatures at neutrino oscillation experiments. Thus, type II radiative neutrino mass model can have sub-eV sterile neutrino, keV sterile neutrino as well as a stable cold dark matter candidate simultaneously. 

In the model with radiative Dirac neutrino mass and tree level right handed neutrino mass, one can have light neutrinos even for light right handed neutrino masses $M_R$. This is due to the one-loop suppression in Dirac neutrino mass term $M_D$ which appears in the light neutrino mass formula for type I seesaw as
$$ m_{\nu} = -M_D M^{-1}_R M^T_D $$
Thus, even for GeV scale right handed neutrinos, one can have sub eV light neutrino mass without unnatural fine tuning in the Dirac neutrino masses. Such GeV scale right handed neutrinos can have interesting implications for experiments like $0\nu \beta \beta$ and colliders, to be discussed below.
\begin{figure}[!h]
\centering
\epsfig{file=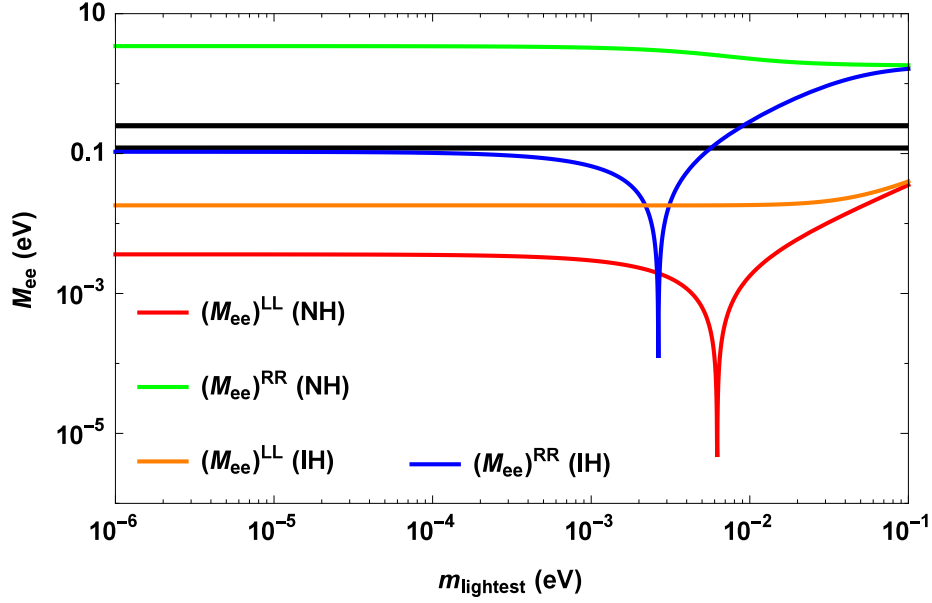,width=0.75\textwidth,clip=}
\caption{Light and heavy neutrino contributions to $0\nu \beta \beta$ for $M_{W_R} = 3$ TeV, lightest right handed neutrino mass $\sim 1$ GeV with Majorana CP phases $\alpha = \beta = \pi/2$. The horizontal black lines correspond to the upper bound $M_{ee}<$ (0.12 - 0.25) eV \cite{kamland_zen} at $90\%$ C.L.}
\label{fig3}
\end{figure}

\begin{figure}[!h]
\centering
\epsfig{file=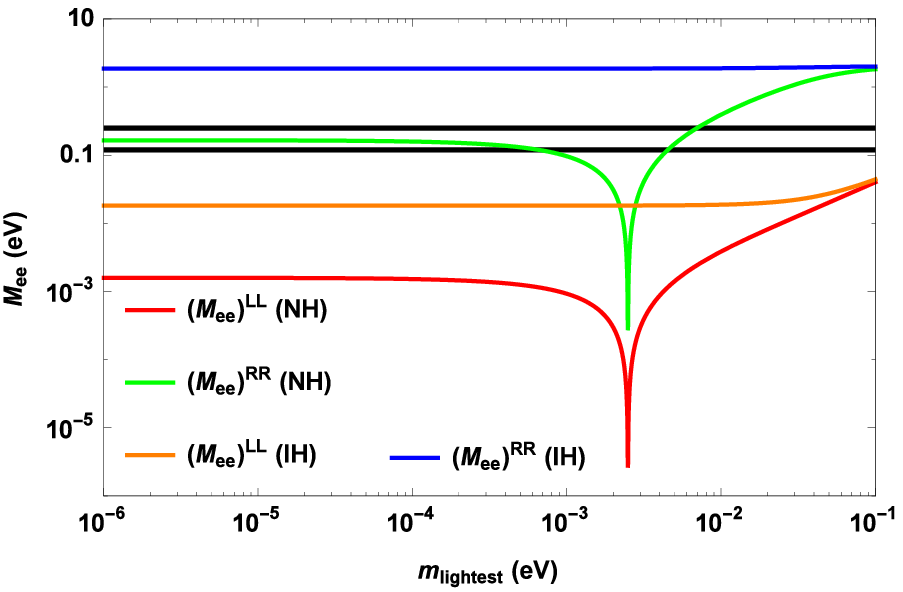,width=0.75\textwidth,clip=}
\caption{Light and heavy neutrino contributions to $0\nu \beta \beta$ for $M_{W_R} = 3$ TeV, heaviest right handed neutrino mass $\sim 10$ MeV with Majorana CP phases $\alpha = \pi/2, \beta = \pi$. The horizontal black lines correspond to the upper bound $M_{ee}<$ (0.12 - 0.25) eV \cite{kamland_zen} at $90\%$ C.L.}
\label{fig31}
\end{figure}

\begin{figure}[!h]
\centering
\begin{tabular}{cc}
\epsfig{file=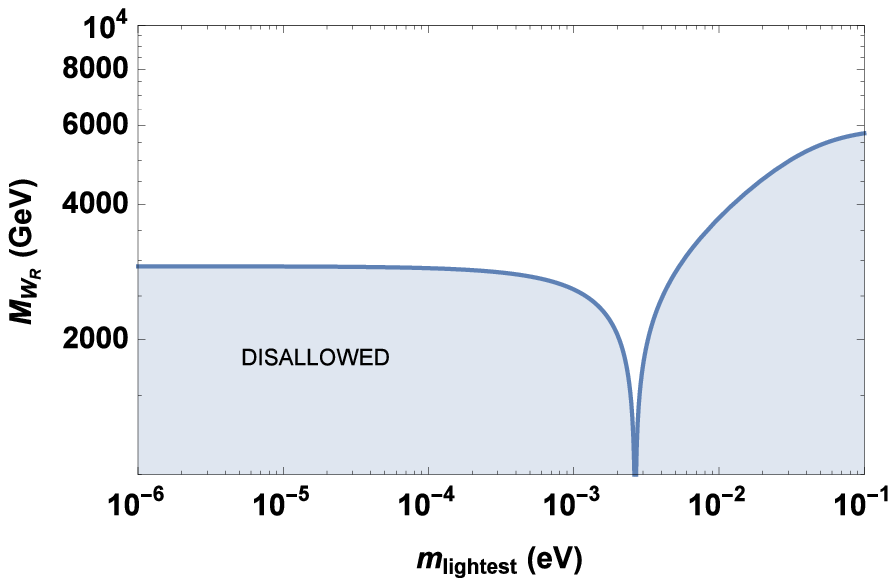,width=0.50\textwidth,clip=} & 
\epsfig{file=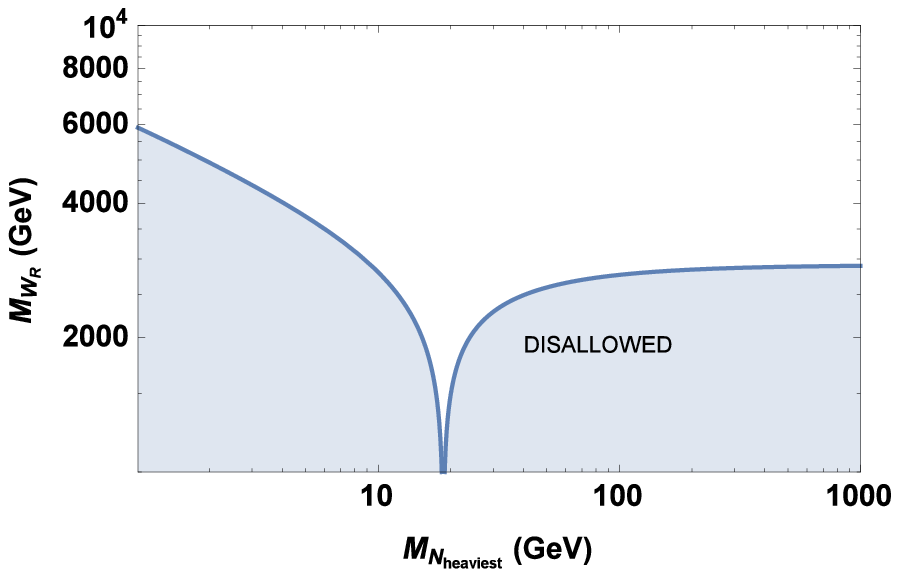,width=0.50\textwidth,clip=}
\end{tabular}
\caption{Constraints on $M_{W_R}-m_{\text{lightest}}$ and $M_{W_R}-M_{N_\text{heaviest}}$ from experimental constraints on $0\nu \beta \beta$ lifetime. Both the panels corresponds to inverted hierarchy with Majorana CP phases $\alpha = \beta = \pi/2$ and lightest right handed neutrino mass $\sim 1$ GeV.}
\label{fig4}
\end{figure}

\begin{figure}[!h]
\centering
\begin{tabular}{cc}
\epsfig{file=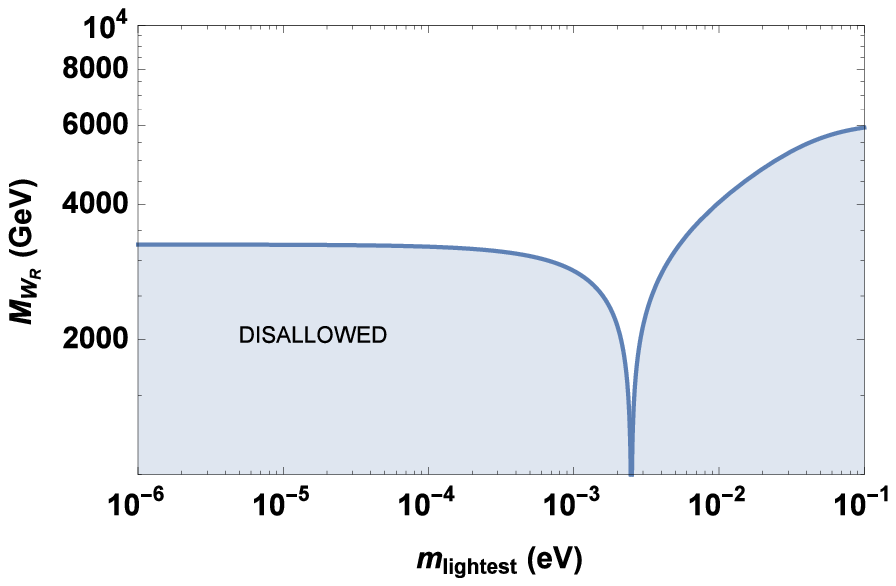,width=0.50\textwidth,clip=} & 
\epsfig{file=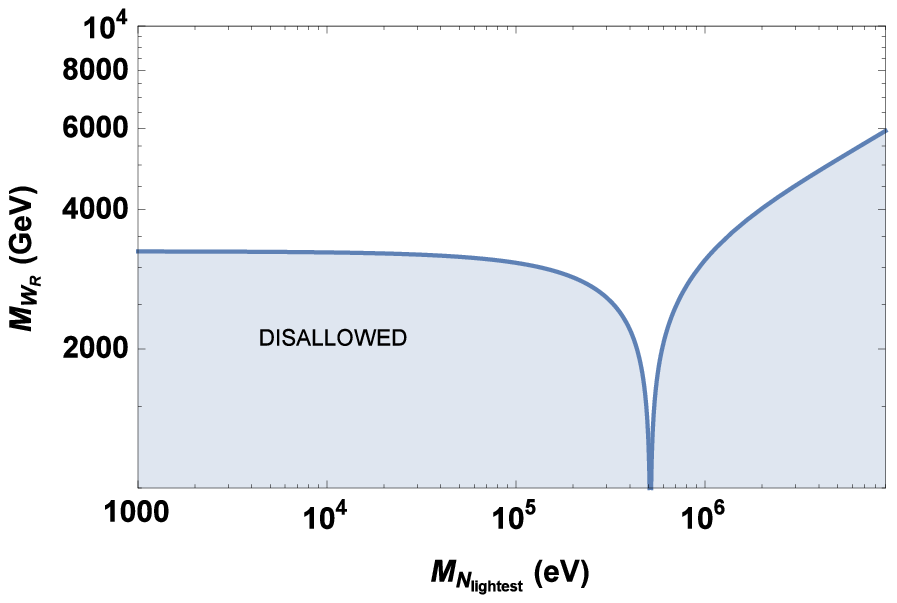,width=0.50\textwidth,clip=}
\end{tabular}
\caption{Constraints on $M_{W_R}-m_{\text{lightest}}$ and $M_{W_R}-M_{N_\text{lightest}}$ from experimental constraints on $0\nu \beta \beta$ lifetime. Both the panels corresponds to normal hierarchy with Majorana CP phases $\alpha = \pi/2, \beta = \pi$ and heaviest right handed neutrino mass $\sim 10$ MeV.}
\label{fig41}
\end{figure}

\section{Implications for $0\nu \beta \beta$}
\label{sec5}
Due to the presence of several new vector, scalar and fermionic particles in LRSM, one can have various new physics contribution to neutrinoless double beta decay $(0 \nu \beta \beta)$. The relevant Feynman diagrams and the calculation details can be found in several earlier works \cite{ndbd00} and references therein. Apart from the SM contribution mediated by light neutrinos and $W_L$ bosons, there can be contributions from (a) Right handed neutrino and $W_R$, (b) Mixing between heavy and light neutrinos (c) $W_L - W_R$ mixing and (d) Triplet scalars in type II seesaw scenarios. Here we consider only the right handed neutrino with $W_R$ contribution as new physics and ignore the active-sterile, $W_L-W_R$ mixing and triplet scalar contributions.

The amplitude of the light neutrino contribution is given by 
\begin{equation}
A_{\nu L L} \propto G^2_F \sum_i \frac{m_i U^2_{ei}}{p^2} 
\end{equation}
where $p \sim 100$ MeV is the average momentum exchange of the process, $m_i (i=1,2,3)$ are the masses of three light neutrinos and $U$ is the standard Pontecorvo-Maki-Nakagawa-Sakata (PMNS) leptonic mixing matrix. The contribution from the heavy neutrino and $W_R$ exchange can be written in the limit $M_i \gg p$ as 
\begin{equation}
A_{NRR} \propto G^2_F \left ( \frac{M_{W_L}}{M_{W_R}} \right )^4 \sum_i \frac{V^{*2}_{ei}}{M_i} 
\end{equation}
where $V$ is the heavy neutrino mixing matrix and $M_i (i=1,2,3)$ are the masses of heavy right handed neutrinos. Thus the total contribution to $(0 \nu \beta \beta)$ process is
\begin{equation}
\frac{\Gamma_{\text{NDBD}}}{\text{ln}2} = G \frac{\lvert \mathcal{M}_\nu \rvert^2}{m^2_e}\left( \big \lvert  M^{LL}_{ee} \big \rvert^2+ \big \lvert M^{RR}_{ee} \big \rvert^2 \right)
\label{eq:ndbd2}
\end{equation}
where 
$$ M^{LL}_{ee} = \sum_i U^2_{Lei}m_i, \;\;\; M^{RR}_{ee}=p^2 \frac{M^4_{W_L}}{M^4_{W_R}} \sum_i\frac{V^{*2}_{ei}}{M_i}$$
are effective neutrino masses corresponding to light neutrino and heavy neutrino exchanges respectively. Here, $\mathcal{M}_\nu$ is the nuclear matrix element.

Now, for LRSM with universal seesaw, the light and heavy neutrino mass matrices are related by a factor $x=\frac{v^2_R}{v^2_{EW}}$ where $v_{EW} = 246$ GeV. Similarly in the radiative type II seesaw model, they are related by a factor $x=\frac{v_R}{v_L}$ which is also true for tree level type II seesaw model. Since the heavy and light neutrino mass matrix is similar upto a factor, we can consider $U=V$ in the above expressions. Also the mass eigenvalues will be related by the same numerical factor $x$. Thus, the new physics contribution can be written as 
$$ M^{RR}_{ee}=p^2 \frac{M^4_{W_L}}{M^4_{W_R}} \sum_i\frac{U^{*2}_{ei}}{ x m_i}$$
Using the standard parameterisation of PMNS mixing matrix in terms of three angles and three CP phases (one Dirac and two Majorana phases), one can write down the light neutrino contribution as
\begin{equation}
M^{LL}_{ee}=m_1c^2_{12}c^2_{13}+m_2s^2_{12}c^2_{13}e^{2i\alpha}+m_3 s^2_{13}e^{2i\beta} 
\end{equation}
where $c_{ij} = \cos{\theta_{ij}}, \; s_{ij} = \sin{\theta_{ij}}$, $\theta_{ij}$ being the mixing angles and $\alpha, \beta$ are Majorana CP phases. Fixing the lightest right handed neutrino mass to be 1 GeV and hence $x = 10^9/m_{\text{lightest}}$, we calculate the heavy neutrino contribution to the effective neutrino mass. The contribution is found to be minimal for Majorana CP phases $\alpha = \beta = \pi/2$ as shown in figure \ref{fig3}. It can be seen from the figure that such a scenario with $M_{N_{\text{lightest}}} = 1$ GeV, $M_{W_R}=3$ TeV and normal hierarchy of light neutrinos is ruled out from existing experimental constraint. For inverted hierarchy also, the region for $m_{\text{lightest}} \geq 0.01$ eV is ruled out whereas the $m_{\text{lightest}} \leq 0.0001$ eV region  contribution almost coincides with the experimental upper bound. Choosing different values of Majorana phases rules out even more parameter space.

Similarly, we calculate the right handed neutrino contribution for masses $\leq 10$ MeV. In such low mass regime, the right handed neutrino contribution has a similar expression as standard light neutrino contribution upto a scaling factor of gauge boson mass ratio. The right handed neutrino contribution is found to be minimal for Majorana CP phases $\alpha = \pi/2, \beta = \pi$ as seen from figure \ref{fig31}. It can be seen that inverted hierarchy of light neutrino spectrum is already ruled out for such a case with $M_{W_R}=3$ TeV. The normal hierarchical scenario survives only in a narrow region of parameter space around $m_{\text{lightest}} \approx 0.0025$ eV. After showing the allowed parameter space for a specific choice of $M_{W_R}$, we show the allowed regions of $M_{W_R}-m_{\text{lightest}}$ and $M_{W_R}-M_{N_\text{heaviest}} (M_{W_R}-M_{N_\text{lightest}})$ parameter space in figure \ref{fig4} and \ref{fig41}, by incorporating the experimental upper bound on $0\nu \beta \beta$ amplitude. As seen from figure \ref{fig41}, one can still have keV scale right handed neutrino along with a TeV scale $W_R$ after incorporating experimental constraints on $0\nu \beta \beta$ lifetime.

\section{Collider Signatures}
\label{sec6}
Depending on the origin of light neutrino masses, there can be different collider signatures for the new physics sector responsible for it. If neutrinos are Dirac fermions such that $\nu_L, \nu_R$ form a Dirac fermion with eV scale Dirac mass, then one can not have on-shell production of same sign dilepton plus dijet signatures mediated by $W_R$ bosons. Such a process was proposed \cite{keungsenja} to be a clean signature of right handed gauge bosons with heavy Majorana neutrinos. In these scenarios $W_R$ boson can decay into $e_R, \nu_R$ final states in a way similar to $W_L$ decay into $e_L, \nu_L$. Also, due to the presence of scalar doublet dark matter candidate, $W_R$ can also have decay into dijet plus missing energy through $W^{\pm}_R \rightarrow  \eta^{\pm}_R, \eta^0_R \rightarrow j j \eta^0_R \eta^0_R$ or $W^{\pm}_R \rightarrow  W^{*\pm}_R, \eta^0_R \eta^0_R \rightarrow j j \eta^0_R \eta^0_R$. Here $\eta^0_R$ in final states can give rise to missing energy, if it is the lightest neutral particle stabilised by the remnant $Z_2$ symmetry.

However, if neutrinos are Majorana fermions and heavy neutrinos are in the GeV-TeV range with TeV scale $W_R$, they can be constrained by LHC data from the search of dilepton plus dijet. For even lower masses of right handed neutrinos say, in the keV regime, LHC sensitivity may not be enough and one has to go for alternate probes like neutrinoless double beta decay, discussed above. Apart from heavy neutrino mediated dilepton plus dijet final states, one can also have other interesting collider signatures due to the presence of additional scalars. The scalar sector of MLRSM and some of its minimal variants have been extensively studied by several groups \cite{collider1}. In the model with radiative type II seesaw, one can have light doubly charged scalars similar to MLRSM. However, their collider signatures can be entirely different. The present LHC lower bounds on doubly charged scalar masses (465-550 GeV for $M_{\delta^{++}_L}$, 370-435 GeV for $M_{\delta^{++}_R}$) \cite{H++LHC} are derived by considering $100\%$ branching ratio into same sign lepton pairs. In the radiative type II seesaw model discussed above, the doubly charged scalars do not have tree level coupling to the leptons and hence their decay into lepton pairs will be suppressed. Thus, the above bounds are not applicable and one can have lighter doubly charged boson as well. In the model discussed here, one can have $\delta^{++}_L \rightarrow W^+_L W^+_L $ or $\delta^{++}_L \rightarrow \eta^+_L \eta^+_L \rightarrow W^+_L W^+_L \eta^0_L \eta^0_L$ with the $W_L$ boson finally decaying into two jets or leptons. Considering leptonic final states only, both the decay modes can contribute to same sign dilepton plus missing energy signatures. Similar conclusions can be made for right handed doubly charged scalar $\delta^{++}_R$ as well.

In the model with radiative type I seesaw, the doubly charged scalars have tree level coupling to leptons. However, the neutral component of the left handed triplet does not acquire any vev. Therefore the decay channel $\delta^{++}_L \rightarrow W^+_L W^+_L $ is absent. However $\Delta_{L,R}$ still couples to scalar doublets $\eta_{L,R}$ and hence both $\delta^{++}_L \rightarrow l^+ l^+$ and $\delta^{++}_L \rightarrow \eta^+_L \eta^+_L \rightarrow W^+_L W^+_L \eta^0_L \eta^0_L$ are possible in such a scenario. Same conclusion can be made for right handed triplet scalar as well. 

\section{Cosmological Implications}
\label{sec7}
\subsection{Light Sterile Neutrinos in Cosmology}
Cosmological implications for sterile neutrinos with masses from eV scale to GeV-TeV scale are well described in the recent review \cite{whitekev}. Here we focus only on the eV-keV range as the above analysis was primarily focussed on generating sterile neutrino masses in this range. The left-right symmetric model with light sterile neutrinos in eV to keV scale along with a cold dark matter candidate as discussed above can have interesting implications for cosmology. If one or more of the right handed neutrinos have masses in the eV range, then they can contribute to the total radiation content of the Universe. Such extra relativistic degrees of freedom may affect the big bang nucleosynthesis (BBN) predictions as well as cause changes in the cosmic microwave background (CMB) spectrum. The latest cosmology data from Planck experiment \cite{Planck15} constrains the the effective relativistic degrees of freedom $N_{\text{eff}} < 3.7$. 

In the model where the light neutrinos are Dirac, the right handed components contribute extra degrees of freedom to $N_{\text{eff}}$. As shown in \cite{nuRdecoupling}, this can evade the BBN bound if the right handed neutrinos decouple much earlier that is $T^D_{\nu_R} > T^D_{\nu_L}$. The $\nu_R$ contribution to $N_{\text{eff}}$ gets diluted due to the decrease in effective relativistic degrees of freedom $g_*$ as the Universe cools down from $T^D_{\nu_R}$ to $T^D_{\nu_L}$:
$$ N_{\text{eff}} \approx 3 + 3 \left (\frac{g_*(T^D_{\nu_L})}{g_*(T^D_{\nu_R})} \right )^{\frac{4}{3}} $$
Since $g_*(T^D_{\nu_L}) = 10.75$ for the relativistic degrees of freedom in SM at $T=T^D_{\nu_L} \approx 1$ MeV, the right handed neutrinos can evade the Planck bound if $g_*(T^D_{\nu_R}) > 32.02$. This implies that the right handed neutrinos should decouple before the QCD phase transition temperature 200-400 MeV. The decoupling temperature of right handed neutrinos can be calculated by following the same procedure for left handed neutrinos and replacing the $W_L$ mass with $W_R$. In terms of $T^D_{\nu_L}$, it can be written as
$$ T^D_{\nu_R} \approx (g_*(T^D_{\nu_R})^{1/6} \left ( \frac{M_{W_R}}{M_{W_L}} \right )^{4/3}  T^D_{\nu_L} $$
Demanding $T^D_{\nu_R} > 400$ MeV and taking $T^D_{\nu_L} \approx 1$ MeV, we can arrive at the bound on $M_{W_R}$ as 
\begin{equation}
M_{W_R} > 4.3 \; \text{TeV}
\end{equation}
where we have considered $g_*(T^D_{\nu_R}) \approx 60$, which is the number of relativistic degrees of freedom just above the QCD phase transition temperature. The bound on $W_R$ mass from cosmology in this scenario is stronger than the direct search limits discussed earlier. Similar bounds will also be applicable in those models where active neutrinos are Majorana but one or two of the sterile or right handed neutrinos have masses in the eV scale.

Even if the light right handed neutrinos decouple before the QCD temperature, they can again be produced at lower temperatures through active-sterile oscillations. This can again be in tension with the Planck bound on the relativistic degrees of freedom. In the scenarios discussed above, such a situation can occur for the model with particle content shown in table \ref{tab:data2}. For the particle content in table \ref{tab:data2}, both active-active, sterile-sterile and active-sterile mass terms can appear at one loop level. Similarly, in the LRSM with universal seesaw also, one can have sizeable active-sterile neutrino mixing by suitably choosing $M^D_N, M^M_N$ in equation \eqref{nonunitary}. Due to large active-sterile mixing, flavour oscillations will produce the sterile neutrinos at low temperatures making such a scenario highly disfavoured from cosmology. Additional new physics is required to suppress such late production of sterile neutrinos through flavour oscillations \cite{kopp}.

The dark matter phenomenology can be very rich if we have a keV scale sterile neutrino along with a stable dark matter candidate with electroweak scale mass or above. As discussed before, such a possibility exists in the radiative type II seesaw model if we want to keep the right handed gauge bosons near the TeV corner. In LRSM with universal seesaw, the symmetry breaking scale has to pushed high in order to have keV right handed neutrinos whereas in the radiative type I seesaw model, unnatural fine tuning is required to generate keV scale right handed neutrino masses, similar to the MLRSM. As discussed in \cite{whitekev, wdmlr1, wdmlr2} such keV sterile neutrinos with gauge interactions, remain in thermal equilibrium in the early Universe and remain like a thermal relic after their freeze-out. Typically, for TeV scale masses of these additional gauge bosons, these sterile neutrinos decouple at a low temperature, leading to overproduction. Such an overproduction can be avoided by late time entropy release from the decay of a heavier particle into the SM particles \cite{entropy1, entropy2} after the freeze-out of keV sterile neutrino. Using these prescriptions, the authors of \cite{wdmlr1} showed that the right handed gauge bosons have to be heavier than 10 TeV for correct abundance of keV sterile neutrino dark matter in MLRSM. Later, a more careful study \cite{wdmlr2} found a tiny window near $W_R$ mass of 5 TeV for which a keV sterile neutrino dark matter of mass $0.5$ keV can give the correct relic abundance. Similar analysis can also be performed within the models discussed in this work. Similarly, the cold dark matter analysis can also be performed in the usual way of calculating relic abundance after freeze-out. Analysis of scalar doublet dark matter in LRSM was recently done by \cite{heeck1}. However, due to the mixed dark matter scenario, a careful detailed calculation is required, as the late time entropy release to suppress the WDM abundance will also affect the abundance of CDM. Such a mixed dark matter model can be very interesting from astrophysical structure point of view, as it may provide a way to solve the small scale structure problem \cite{kamada}. This can also be interesting from dark matter experiments point of view as the cold component can provide signatures at gamma ray experiments \cite{GCfeb} whereas the warm component with mass around a few keV can be responsible for the anomalies at X-ray telescopes \cite{Xray1, Xray2, Xray3}.

\subsection{Formation of Domain Walls}
The models studied in this work have additional discrete symmetries like $Z_4$ which gets spontaneously broken. Such spontaneous breaking of exact discrete symmetries may have serious cosmological implications as they lead to the formation of domain walls (DW). These domain walls, if stable on cosmological time scales, will be in conflict with the observed Universe \cite{Kibble:1980mv,Hindmarsh:1994re}. Generic left-right symmetric models suffer from the problem of DW due to the spontaneous breaking of discrete left-right symmetry, known as the D-parity. This is a $Z_2$ symmetry which relates the couplings in the left and right sectors in the LRSM. Several works \cite{Mishra:2009mk} pointed it out and also studied a possible way to get rid of these walls. If these walls appear before cosmic inflation, then their density in the present Universe will be too diluted to be of any relevance. But even if we do not assume anything about the scale of inflation, then also one can address the issue of DW by adopting the mechanism suggested by \cite{Rai:1992xw,Lew:1993yt}. The authors in these works considered higher dimensional Planck scale suppressed operators to be the source of domain wall instability which make them disappear. As observed by the authors of \cite{Mishra:2009mk}, the implementation of this mechanism as a solution to the DW problem in LRSM typically puts an upper bound on the scale of gauge symmetry breaking. Implementing the same mechanism in all different versions of LRSM discussed in this work is beyond the scope of this present work. Since this is expected to put only an upper bound on the scale of symmetry breaking, the low energy phenomenology discussed in this work are not going to be affected. However, such an implementation of DW disappearance mechanism should also take care of the longevity of the dark matter candidate in the model. A detailed study of these possibilities is left for a future work.
\section{Conclusion}
\label{sec8}
We have presented a class of left-right symmetric models without the conventional Higgs bidoublet that allows the possibility of having light sterile neutrinos along with a cold dark matter candidate stabilised by additional discrete symmetries. The charged leptons acquire masses through a universal seesaw mechanism due to the presence of additional vector like fermions. The neutrinos can either acquire masses through the same universal seesaw or through one loop radiative corrections, if suitable particles are added with non-trivial transformations under additional symmetries. The same additional symmetry can also stabilise one of the particles going inside the loop in neutrino mass diagram, resulting in a stable dark matter candidate. Depending on the additional particles and symmetries, the light neutrinos can be either Dirac or Majorana. In case of Dirac light neutrinos, the dark sector is composed of only cold dark matter component. However, in case of Majorana light neutrinos, the one or more of the right handed neutrinos can have mass in the eV-keV range without any unnatural fine tuning of the Yukawa couplings. A keV sterile neutrino can be a warm dark matter candidate giving rise to a mixed dark matter scenario along with the cold component. Such light right handed neutrinos can have interesting implications for neutrinoless double beta decay and collider experiments, constraining the right handed gauge boson masses. Cosmology constraints on the number of relativistic degrees of freedom also constraints the right handed gauge boson masses to be heavier than a few TeV so that the right handed neutrinos get decoupled before the QCD phase transition temperature. A detailed calculation of the relic abundance of such mixed dark matter scenario is worth investigating and we leave it for a future work. We also briefly comment upon the possible ways to get rid of domain walls that are formed due to the spontaneous breaking of discrete symmetries in the models.
\begin{acknowledgments}
The author would like to express a special thanks to the Mainz Institute for Theoretical Physics (MITP) and the Munich Institute for Astro and Particle Physics (MIAPP) for hospitality and support while some part of this work was completed.
\end{acknowledgments}

\end{document}